\documentclass[%
 reprint,
 amsmath,amssymb,
 aps,prd,
floatfix,
]{revtex4-2}

\usepackage{graphicx}
\usepackage{dcolumn}
\usepackage{bm}
\usepackage{xcolor}
\usepackage{float}

\usepackage{hyperref}

\begin{document}



\title{  Precision study of the continuum SU(3) Yang-Mills theory: how to use parallel tempering to improve on supercritical slowing down for first order phase transitions }

\author{S. Bors\'anyi} \email{borsanyi@uni-wuppertal.de}
\affiliation{University of Wuppertal, Department of Physics, Wuppertal D-42119, Germany}

\author{Z. Fodor}
\affiliation{Pennsylvania State University, Department of Physics, University Park, PA 16802, USA}
\affiliation{University of Wuppertal, Department of Physics, Wuppertal D-42119, Germany}
\affiliation{Inst.  for Theoretical Physics, ELTE E\"otv\"os Lor\' and University, P\'azm\'any P. s\'et\'any 1/A, H-1117 Budapest, Hungary}
\affiliation{J\"ulich Supercomputing Centre, Forschungszentrum J\"ulich, D-52425 J\"ulich, Germany}

\author{D. A. Godzieba}
\affiliation{Pennsylvania State University, Department of Physics, University Park, PA 16802, USA}

\author{R. Kara}
\affiliation{University of Wuppertal, Department of Physics, Wuppertal D-42119, Germany}

\author{P. Parotto}
\affiliation{Pennsylvania State University, Department of Physics, University Park, PA 16802, USA}

\author{D. Sexty}
\affiliation{University of Graz, Institute for Physics, A-8010 Graz, Austria}

\date{\today}


\begin{abstract}
We perform large scale simulations to characterize the transition in quenched QCD. It is shown by a rigorous finite size scaling that the transition is of first order. After this qualitative feature quantitative results are obtained with unprecedented precision: we calculate the transition temperature $w_0T_c$=0.25384(23) -- which is the first per-mill accurate result in QCD thermodynamics -- and the latent heat $\Delta E/T_c^4$=1.025(21)(27), in both cases carrying out controlled continuum and infinite volume extrapolations. As it is well known the cost of lattice simulations explodes in the vicinity of phase transitions, a phenomenon called critical slowing down for second order phase transitions and supercritical slowing down for first order phase transitions. We show that a generalization of the parallel tempering algorithm of Marinari and Parisi [Europhys. Lett. 19, 451 (1992)] originally for spin systems can efficiently overcome these difficulties even if the transition is of first order, like in the case of QCD without quarks, or with very heavy quarks. We also report on our investigations on the autocorrelation times and other details.
\end{abstract}

\maketitle



\section{\label{sec:Introduction}Introduction} 

There is no doubt in the theoretical physics community that the SU(3)
Yang-Mills theory, or QCD without quarks, features a first order phase
transition. The transition takes place at a temperature around 290~MeV,
the precise value strongly depending on the choice of the variable to
set the scale. It separates the
cold confining phase from a hot phase with spontaneously broken center symmetry.

In the context of full QCD this transition was expected to happen
as a consequence of Hagedorn's exponential spectrum \cite{Hagedorn:1965st},
and a first phase diagram was drawn in Ref.~\cite{Cabibbo:1975ig}.
Later, the necessity of a transition was derived from the underlying quantum
field theory and linked to the center of the gauge group \cite{Polyakov:1978vu}.
In contrast to the well known crossover transition in full QCD \cite{Aoki:2006we}, the order of the transition between the confining and weak-coupling phases was argued to be first order in the quarkless SU(3) theory
\cite{Svetitsky:1982gs,Yaffe:1982qf}. The main argument relies on
the $Z(3)$ symmetry of the three-dimensional effective theory, and that no
stable renormalization group fixed point is known with that symmetry.
Such an effective theory has already been studied extensively in the context of
the Potts model \cite{Wu:1982ra}.

Our main motivation to study the transition of the SU(3) Yang-Mills theory
is the exploration of the so-called ``Columbia plot'', a diagram showing the
nature of the QCD transition as a function of the light and strange quark masses
$ m_u=m_d=m_l, m_s$.
Because of the absence of an exact $Z(3)$ symmetry for finite quark mass
the transition will turn into crossover below a certain critical mass
\cite{Yaffe:1982qf}. This mass was estimated using Dyson-Schwinger equations
\cite{Fischer:2014vxa} and in effective models \cite{Kashiwa:2012wa}.
In the opposing corner of the Columbia plot, referred to as the ``chiral limit'',
one again expects a first order transition 
for three or more massless flavors, following from the $\epsilon$ expansion
of a linear $\sigma$ model \cite{Pisarski:1983ms}.
This diagram was first studied in lattice QCD by the Columbia group
\cite{Brown:1990ev} and has been a subject of intense research ever since.
In terms of simulation costs the study of the upper right corner is
considerably simpler than working close to the chiral limit, partly
because of the slow convergence of the iterative solvers, partly because
of the more elaborate fermion formulations that the chiral limit requires.
However, simulations are in progress even near the chiral corner, and 
recent results challenge the theoretical assumptions on the phase diagram
\cite{Cuteri:2021ikv}.

But even the upper right corner is non-trivial to simulate. It has been
addressed using reweighting from quenched QCD
\cite{Saito:2011fs,Ejiri:2019csa}, reweighting from a leading order
hopping-term expansion \cite{Kiyohara:2021smr} or a three-dimensional effective
theory for the Polyakov loop \cite{Fromm:2011qi}, or with
direct simulations \cite{Cuteri:2020yke,Kara:2021btt}. While these results are
promising and appear to be consistent with matrix models \cite{Kashiwa:2012wa},
a continuum extrapolation could not be achieved, yet.

The main difficulty in pinpointing the critical mass is not simply coming from
the presence of quarks, as the iterative solvers converge very quickly in these
studies. The problem is more general: the critical slowing down becomes severe as
the critical mass is approached. But even at a safe distance from the critical mass,
simulations on the first order side of the phase transition line can be difficult
\cite{Cuteri:2020yke}, the exploration of both phases with the proper
statistical weight turns out to be extremely challenging.

Even in the quarkless SU(3) Yang-Mills theory, the
clear evidence of the expected
first order nature of the transition was quite challenging to obtain.
Followed by the pioneering papers 
\cite{Kuti:1980gh,McLerran:1980pk} 
to observe the transition in a gauge theory
Kajantie et. al. observed a hysteresis in
the Polyakov loop, the first numerical hint for a first order transition
\cite{Kajantie:1981wh}.  The initial two-state signal
\cite{Celik:1983wz,Gottlieb:1985ug} was followed by first estimates of the
latent heat \cite{Kogut:1983mn,Svetitsky:1983bq}. To determine the nature of a
transition from finite-box simulations, a finite size scaling is necessary.
This was done for fixed (coarse) lattice spacing in
Refs.~\cite{Brown:1988qe,Fukugita:1989yb}.  However, not all works agreed with
this: Refs.\cite{Alves:1990pn,Bacilieri:1989ir} proposed a second order
transition.  In fact, the SU(3) transition is weak. In
comparison to other SU(N) theories \cite{Lucini:2002ku}, the latent heat in
SU(3) falls short of the linear $1/N^2$ scaling \cite{Lucini:2005vg}. 

While spin model simulations underwent great progress thanks to
cluster algorithms \cite{Wolff:1988uh}, the gauge theory studies cited here
mostly relied on local update algorithms. In such cases the only tool at hand is the
reweighting method of Ferrenberg and Swendsen \cite{Ferrenberg:1988yz,Ferrenberg:1989ui}, which allows
combination and interpolation between simulation results.
Ref.~\cite{Marinari:1992qd} introduced a tempering algorithm to
circumvent ``freezing simulations'' in spin models. In a nutshell,
the algorithm treats temperature (or any other control parameter) as
a dynamical variable, that walks around in the region of interest.
The strong dependence of the order parameter on the dynamical control
parameter fosters a quick decorrelation. Parallel tempering is
a variant of this idea where a simulation is started for each possible
parameter value. This has been applied succesfully to e.g. various spin models \cite{Swendsen1986replica,hukushima1996exchange}, 
adjoint SU(2) Wilson action modified by a $Z_2$ monopole suppression term
\cite{Burgio:2006xj,Burgio:2006dc}, and to lattice QCD
\cite{Boyd:1997rb,Joo:1998ib,Ilgenfritz:2001jp,Borsanyi:2021gqg},
though in the latter case not for the study of phase transitions.

In this work we advocate the use of parallel tempering in lattice QCD
thermodynamics and demonstrate its virtues by presenting a precise description
of the phase transition in the SU(3) gauge theory, calculating the transition
temperature and the latent heat. 
We do this to lay the foundations of a longer
term work with dynamical fermions \cite{Kara:2021btt}.  
Although there have been many studies on this transition, a complete infinite
volume limit of continuum extrapolated observables has not been calculated, yet.
In Section \ref{sec:tempering} we describe the parallel tempering algorithm and
show the resulting improvement in the autocorrelation times and simulation
precision.  In Section \ref{sec:tc} we present our determination of the critical
couplings for the tree-level Symanzik improved gauge action used in this study,
the continuum extrapolated transition temperature and its infinite
volume limit. We confirm the first order nature of the transition by
performing a finite volume analysis of the continuum extrapolated
susceptibilities in Section \ref{sec:firstorder}. Finally, we calculate the
latent heat and compare this to the latest result in the literature
\cite{Shirogane:2020muc} in Section \ref{sec:latent}, then conclude.


\section{\label{sec:tempering}Tempered simulations of the transition} 

\subsection{The parallel tempering algorithm}

We first describe the parallel tempering algorithm \cite{Swendsen1986replica,Joo:1998ib,marinari1998optimized,earl2005parallel}. Those who are more interested in the final
physics results can skip this chapter.

Suppose we want to simulate some theory at various parameter sets $p_i$
(which can include the gauge coupling $\beta$, fermionic masses, etc.). 
Connected to each parameter set we have a sub-ensemble $\Gamma_i$ containing
configurations (link variables, pseudo-fermion fields, etc.)
and an action $S_i$ dependent on the parameters $p_i$ and on
configurations from $\Gamma_i$.
The parallel tempering simulation considers a Markov chain built on the
direct product of the sub-ensembles' configuration spaces:
$ \Gamma_\textrm{PT}= \prod_{i=1}^N \Gamma_i $.
Our intent is to build a Markov chain that equilibrates to the distribution
\begin{eqnarray}
P^\textrm{eq}_\textrm{PT}[ \{ a_i\}] = \prod_i P^\textrm{eq}_i ( a_i )
= \prod_i {1\over Z_i} e^{-S_i( a_i ) },
\end{eqnarray}
where $a_i$ are configurations from $\Gamma_i$ and 
$Z_i$ is the partition function of sub-ensemble $\Gamma_i$.
The partition function of the whole ensemble is than
calculated as
\begin{eqnarray}
Z_{PT} = \prod_i Z_i.
\end{eqnarray}  

We define two kinds of transitions on $ \Gamma_\textrm{PT}$:
\begin{enumerate}
\item {\bf Transitions within a sub-ensemble:} We can use any Markovian
  updating procedure (HMC, local Metropolis update, etc.) on the
  sub-ensemble $\Gamma_i$ using the action $S_i$.
\item  {\bf Swapping update of two sub-ensembles:}  This step mixes
  the phase space of $\Gamma_i$ and $\Gamma_j$. We propose to swap
  configurations $a \in \Gamma_i $ and $b \in \Gamma_j$ with probability
  $P_s(i,j)$.
\end{enumerate}  

To satisfy detailed balance of the swapping updates, we must have
\begin{eqnarray}
P_s(i,j) e^{-S_i(a)} e^{-S_j(b)} =   P_s(j,i) e^{-S_i(b)} e^{-S_j(a)} 
\end{eqnarray}  
This is easily satisfied with the usual Metropolis acceptance:
$ P_s(i,j) = \textrm{min} ( 1, e^{-\Delta S } )$, with
\begin{eqnarray}
\Delta S = S_i(b) + S_j(a) - S_i(a) - S_j(b) 
\end{eqnarray}
To reach the desired equilibrium distribution, updates within a sub-ensemble
are essential, they would suffice without any swapping updates (that would be
equivalent to running independent calculations for each parameter set). As
we show below, swapping updates reduce the autocorrelation times
of the simulations immensely. The intuitive understanding behind this
observation is that the swapping updates couple parameter sets with
different autocorrelation times $\tau$.
As configurations do a ``random-walk'' in parameter space, they
decorrelate quickly at the low $\tau$ region, and wander back to the high
$\tau$ region to contribute an independent configuration. This is a more
efficient way to produce independent configurations than running the
Markov chain at a fixed parameter set.

For the parallel tempering to be effective, the acceptance rate
of swapping updates must be carefully controlled, such that the
action distributions of neighboring ensembles have a
substantial overlap.
This can be achieved through the control of the distance of
the parameter sets $p_i$, which, thus, have to move closer to each other
as the physical volume increases.
This is easily satisfied 
if the number of streams in the transition region is kept constant as the
width of the transition region scales with inverse volume.

To locate the phase transition point for a given spatial and temporal lattice size, we perform a $\beta$-scan
to search for the peak of the Polyakov-loop susceptibility or the zero of the third Binder cumulant
(see Sec.~\ref{sec:tc} for details). Typically, we have simulations at 16-256 $\beta$ values, and we use the
parallel tempering
algorithm.  We introduce swapping updates at predefined points in the Markov
chain (typically after 5 sweeps in each sub-ensemble).
This algorithm can be especially efficiently parallelized in our case where the $\beta$ dependence of the
action is simply given as an overall factor of some function of the
link variables.
We set up a number of streams updating a configuration at certain $\beta$ values (typically we use equidistant points).
After 5 sweeps on their configuration, all the streams need to communicate one number (their action) to a
master node which proposes (and accepts or rejects)
several swapping updates for each stream, and afterwards informs each stream which $\beta$ value they ended up at.
This means that the network bandwidth and computational requirements for the swapping updates are negligible, although
some efficiency is lost as synchronization between the streams is required.
Note that this synchronization loss is independent of the time that is needed
for the calculation of one update on the configurations (e.g. it is $<10\%$
if a stream waits on average less then half of an update time after every 5 updates).

\subsection{Explorations with the tempering algorithm}

Below we show results of the parallel tempering method comparing it with the brute force (independent
simulations at each $\beta$ value) and brute force with the multihistogram method.

\begin{figure}
	\centering
	\includegraphics[width=0.88\linewidth]{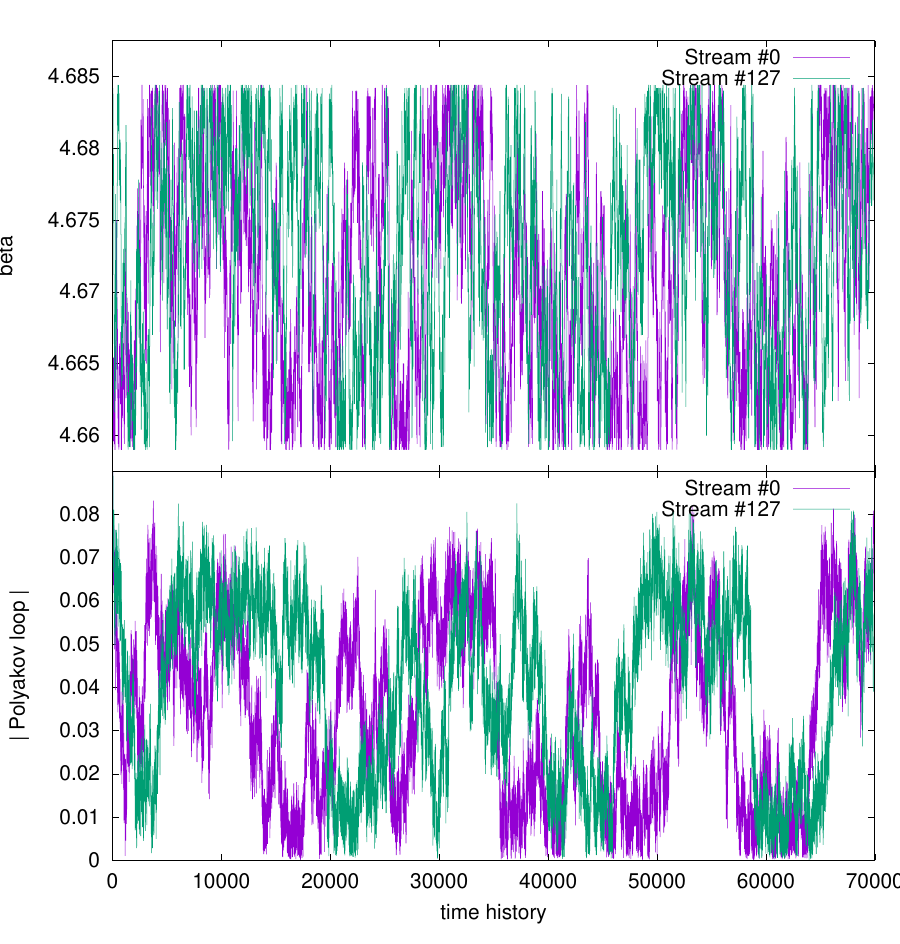}
	\caption{The $\beta$ history (above) and Polyakov loop average (below) of 2 streams
         as a function of the Monte Carlo sweeps
         in a simulation on a $45^3 \times 10$ lattice with 128 streams on
         the range $ 4.659 \le \beta \le 4.6844 $    divided equidistantly.
        }
	\label{temperinghist}
\end{figure}

To set up the simulations we need to choose the
set of $\beta$ values we want to carry out simulations at.
The distance between neighboring ensembles, $\Delta \beta$ needs
to be small enough such that the acceptance of the swapping updates is $O(1)$ (see below for more details).

\begin{figure}
	\centering
	\includegraphics[width=1\linewidth]{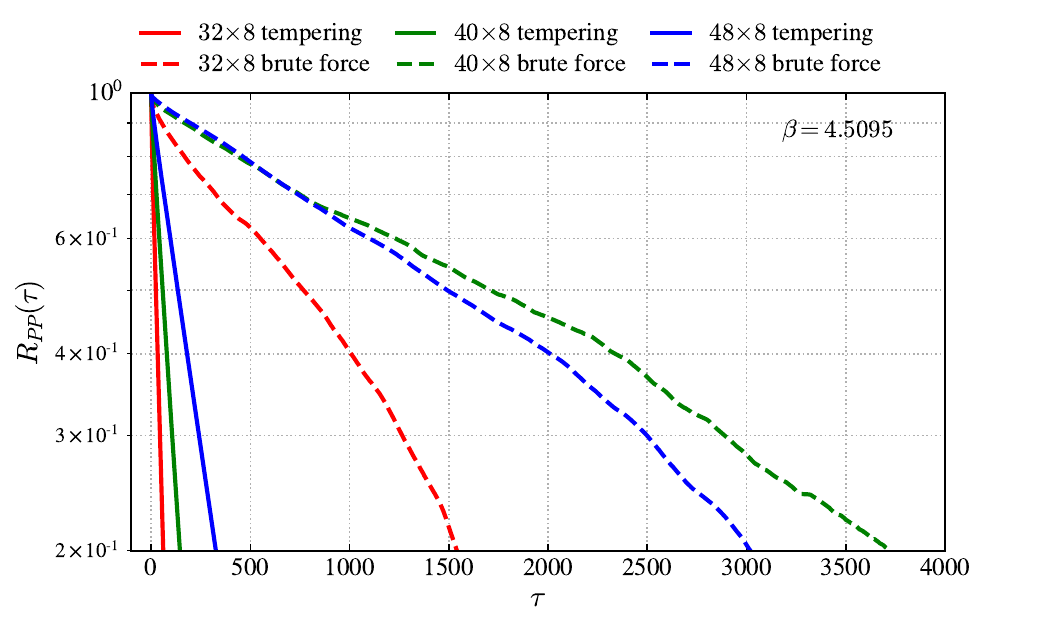}
	\caption{The autocorrelation function of the absolute value of the Polyakov loop at $\beta=4.5095$
        for the lattice sizes $ 32^3\times 8,\ 40^3\times 8,\ 48^3\times 8$, in a parallel tempering simulation (solid lines), using 64 $\beta$ values with spacings $\Delta \beta=0.001 $, and in single simulations (dashed lines).}
	\label{autocorr}
\end{figure}

\begin{figure*}
	\centering
	\includegraphics[trim=50 40 20 0, clip, width=0.8\linewidth]{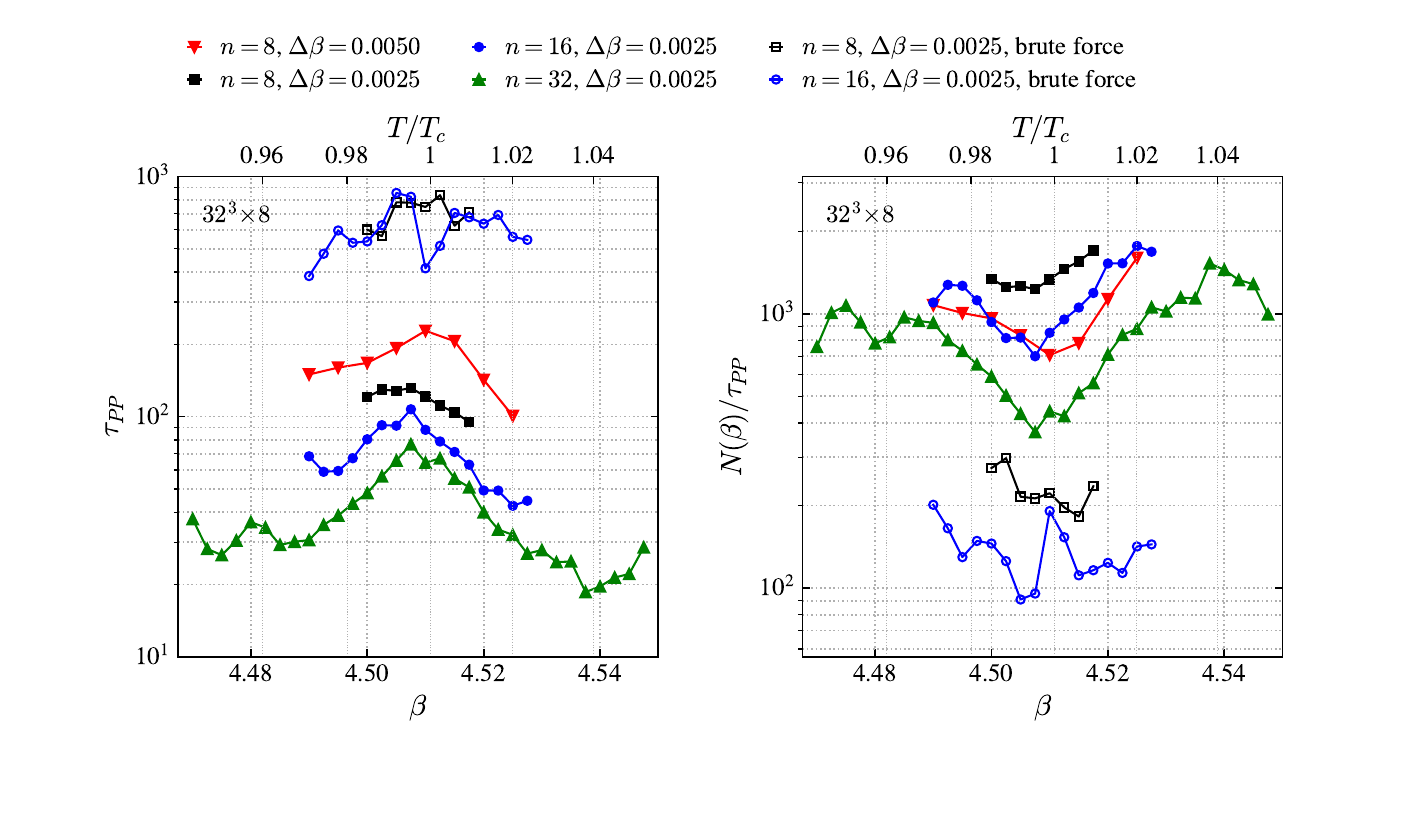}
	\caption{(Left) The autocorrelation time of the Polyakov loop as a function of $\beta$ for simulations on a $32^3{\times}8$ lattice computed using the tempering algorithm (solid points) and using the brute force method (unfilled points) around the transition temperature for $N_t=8$ with equal amounts of computer time. The number of $\beta$ values and their spacing, $\Delta\beta$, are given for each simulation. $\tau_{PP}$ is improved by about an order of magnitude due to the stream-swapping of the tempering algorithm compared to that of the brute force approach. $\tau_{PP}$ also improves greatly as the number of $\beta$ values increases (increasing the number of participating streams at any given $\beta$) and as $\Delta\beta$ decreases (increasing the probability of any one stream being swapped). (Right) The efficiency of each simulation in terms of the number of computed updates divided by the autocorrelation time. }
	\label{autocorrelation time}
\end{figure*}

In Fig.~\ref{temperinghist} we show the $\beta$-history of two streams (out of 128) for a simulation
on a $45^3 \times 10 $ lattice. As one observes the streams follow a trajectory similar to a random walk
on the allowed $\beta$ range and they visit all points for a long enough simulation.
This introduces correlations between ensembles at different $\beta$ values, which we will take into
account later. As also visible on Fig.~\ref{temperinghist}, the value
of the Polyakov loop changes relatively slowly during the history of the stream. If we look
at the history of the Polyakov loop for a given $\beta$, the largest autocorrelation comes from the
instances when the same stream contributes (even if the stream visited other $\beta$ values in the meantime). We
therefore reorder the Monte Carlo chain using the stream ID number
of the contributions (and keep the chronological order among contributions from each stream). This ensures
that the most
correlated contributions will be close to each other
(i.e., this ordering results in the largest autocorrelation for the given Monte-Carlo chain), which
helps to minimize the correlations between blocks in the jackknife analysis.
In Fig.~\ref{autocorr} we show the resulting autocorrelation function for several lattice sizes at
$\beta=4.5095$, compared with autocorrelation functions of brute force simulations.
In the left panel of Fig.~\ref{autocorrelation time} we show the measured autocorrelation times for various simulation setups.
As one observes the autocorrelation is substantially smaller for the parallel tempering simulations,
although at the increased cost of having to simulate at all $\beta$ values. Therefore
a fairer comparison would consider several brute force simulations at all $\beta$ values.
In the right panel of Fig.~\ref{autocorrelation time} we show the number of statistically independent configurations 
created in brute force and in parallel tempering simulations at similar parameters, using
the same amount of computational resources.
We observe that $\tau_{PP}$, the autocorrelation time of the Polyakov loop also improves greatly as the number of $\beta$ values increases in a given $\beta$ range and as $\Delta\beta$ decreases (increasing the acceptance rate of swap updates).
From the red and black points in Fig.~\ref{autocorrelation time}, we see that the number of independent configurations
created by the tempering algorithm increases substantially by simply decreasing $\Delta\beta$. Increasing the density of $\beta$ values, as between the red and blue points, results in nearly the same efficiency, indicating that the decrease in the number of updates (due to having twice as many streams for the same computer time) scales roughly with the decrease in the autocorrelation time.

\begin{figure*}
	\centering
	\includegraphics[width=0.68\linewidth]{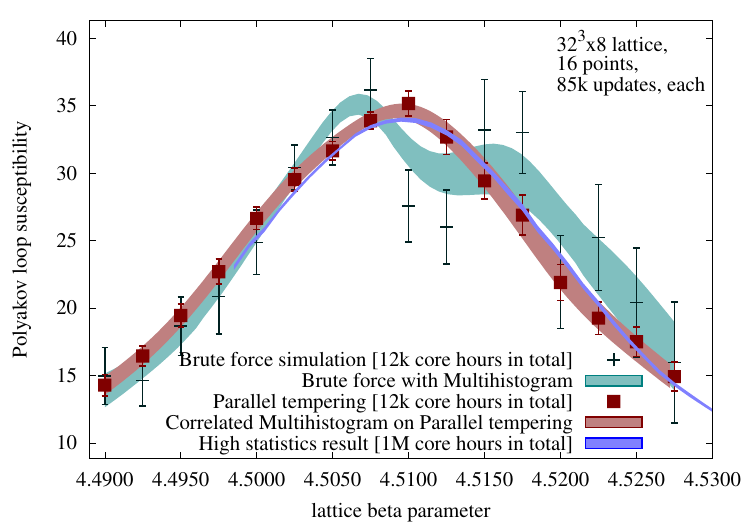}
	\caption{The susceptibility of the Polyakov loop average Eq.~(\ref{eq:susceptibility definition})
          measured on a $32^3\times 8$ lattice with several different algorithms and analyses as indicated,
          using 12k core hours of computer time.
          One local update step consists of three complete sweeps of overrelaxation and one heatbath step at each site and each of the three SU(2) subgroups. On the $32^3\times8$ lattice the full local update completes in $0.64$s
on a 68-core KNL card.
          A high statistics result is also shown for comparison.
        }
      	\label{suscmeas-experiment}
\end{figure*}

Finally, in Fig.~\ref{suscmeas-experiment} we show one of the observables of interest: the Polyakov
loop susceptibility. We show results from parallel tempering and brute force simulations using the same
amount of computational resources. For comparison we also show the ``correct'' result we calculated
using much larger statistics. We observe that the parallel tempering simulations give much smaller statistical errors.
Although the brute force simulations are normally analyzed with the multihistogram
method \cite{Ferrenberg:1989ui}, we see that
the parallel tempered simulations are superior to that as well. Moreover, one can also use the multihistogram
method for the results of the parallel tempering algorithm, for a slight reduction of its errors,
as most of the benefit of information of nearby ensembles is gained already using the swapping updates.
As the parallel tempering introduces correlations between ensembles at different $\beta$s, the
multihistogram analysis becomes slightly more complicated (see in Appendix~\ref{app:corrmulti}). In fact,
the multihistogram algorithm is based on a linear combination of the reweighted results of all the available ensembles
to minimize statistical errors. Therefore, if we use the uncorrelated
multihistogram procedure on the parallel tempered ensembles, we might get slightly larger error bars, as we use a suboptimal linear combination. In practice,
the correlations between ensembles are quite small, neglecting these correlations in the multihistogram
procedure causes a negligible change in the results (much smaller than the statistical errors).

To summarize, we find the usage of the parallel tempering algorithm is very beneficial in these investigations.
A multihistogram analysis helps to further reduce the statistical errors only slightly, as most of the effect
of the swap updates is present already in the single $\beta$ ensembles.


\section{\label{sec:tc}Transition Temperature} 

\begin{figure}
	\centering
	\includegraphics[trim=0 0 0 0, clip, width=\linewidth]{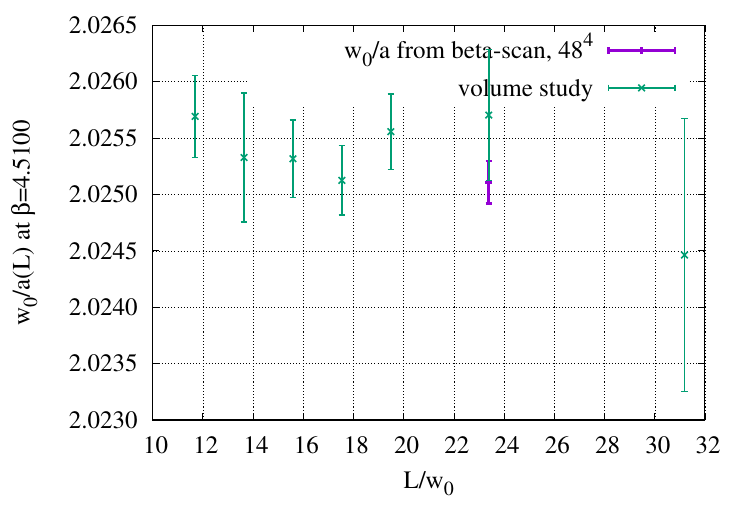}
	\caption{Volume dependence of the $w_0$ scale. $w_0/a$ is computed using fixed bare quark masses that correspond to the physical point, and is shown as a function of the spatial extent of the box in lattice units $L/w_0$. Fluctuations are comparable to the statistical error at each volume in the study; hence, finite-volume effects remain small.}
	\label{fig:w0/a volume dependence}
\end{figure}

\begin{figure*}
	\centering
	\includegraphics[trim=0 0 0 0, clip, width=\linewidth]{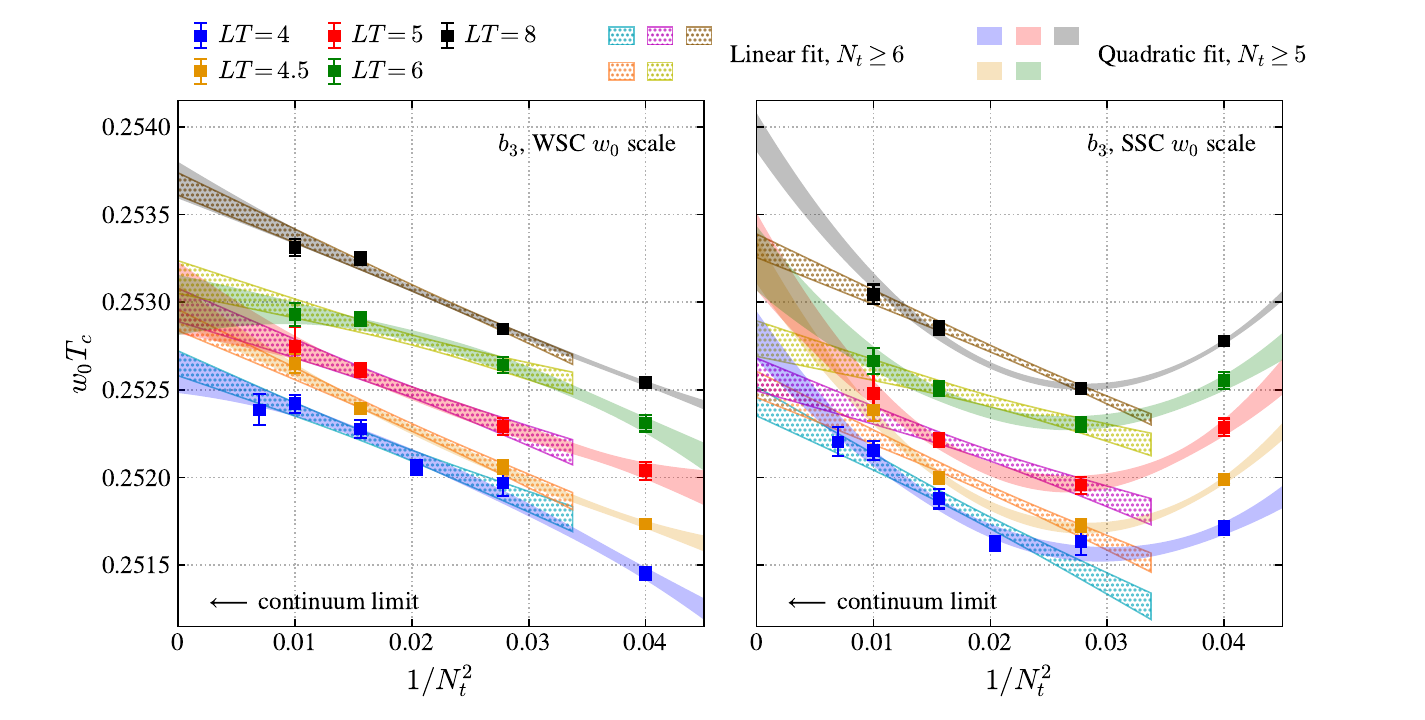}
	\caption{Two example plots that summarize the continuum extrapolation of the transition temperature $w_0 T_c$ computed from the zero-crossing of $b_3$ at five fixed lattice aspect ratios $LT$; the zero-crossing is converted into the quantity $w_0 T_c$ using (Left) the WSC $w_0$ scale and (Right) the SSC $w_0$ scale. Each data point is the median value of $w_0 T_c$ computed for that lattice from the systematic analysis, and the error bars give the combined statistical and systematic error. Error bands are shown for the two kinds of extrapolating fit that are used: a linear fit to lattices with $N_t \geq 6$, and a quadratic fit to lattices with $N_t \geq 5$. The error bands give the combined statistical and systematic error from all similar fits performed in the analysis.}
	\label{fig:w0*T continuum limit}
\end{figure*}

\begin{figure*}
	\centering
	\includegraphics[trim=0 0 0 0, clip, width=0.75\linewidth]{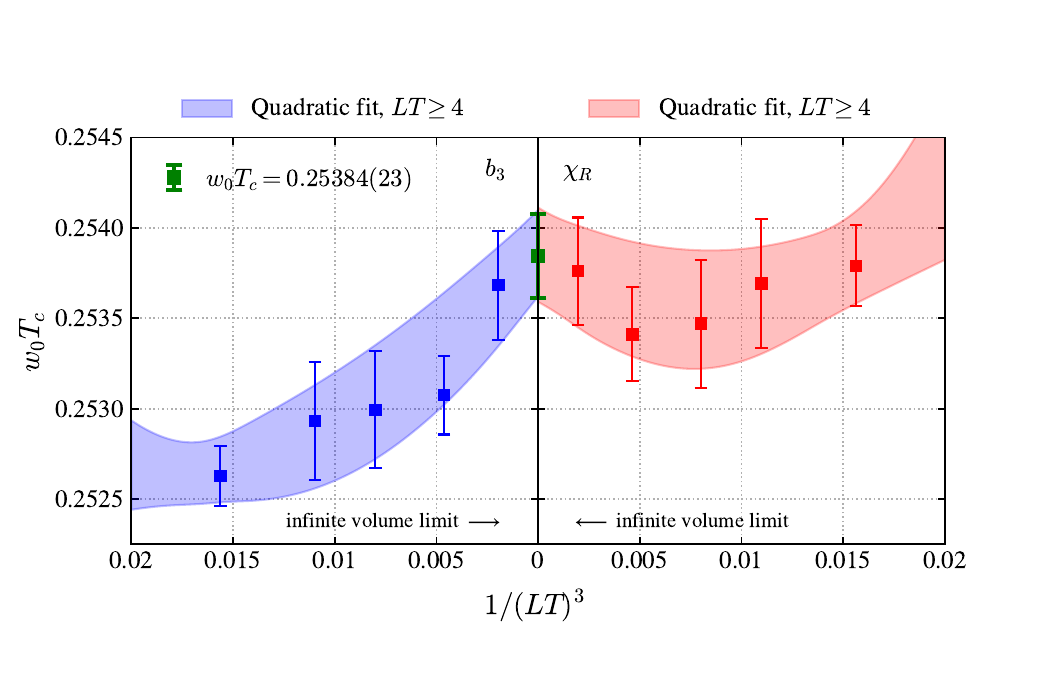}
	\caption{A summary plot of the infinite volume extrapolation of the continuum transition temperature $w_0 T_c$ computed (Left) from the zero-crossing of $b_3$ and (Right) from the peak of $\chi_R$. Each point is the median value of $w_0 T_c$ computed for each aspect ratio $LT$ in the systematic analysis, and the error bars give the combined statistical and systematic error. Error bands are shown for the quadratic fit to $LT \geq 4$. The error bands give the combined statistical and systematic errors from all similar fits in the analysis. The extrapolations from $b_3$ and $\chi_R$ are in agreement, as expected. The final result, $w_0 T_c = 0.25384(23)$, is shown in green.}
	\label{fig:w0*T infinite volume limit}
\end{figure*}

\begin{figure} 
   \hspace*{-8mm}	
   \includegraphics[trim=0 0 0 0, clip, width=1.1\linewidth]{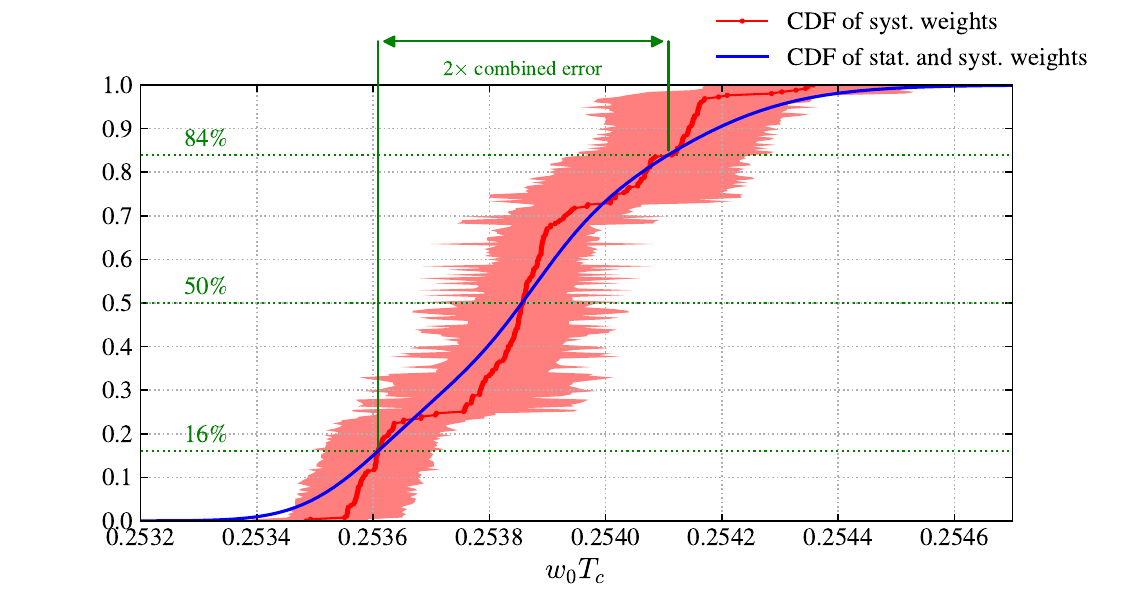}
	\caption{The cumulative distribution function (CDF) of all the results from the systematic analysis of the infinite volume limit of $w_0 T_c$ in the continuum. The red curve shows the central value for each analysis, and the error band shows the statistical error. The blue curve gives the synthesized CDF defined in Eq. (\ref{eq:CDF definition}) combining the statistical and systematic errors. The final result is $w_0 T_c = 0.25384(23)$.}
	\label{fig:w0*T CDF}
\end{figure}

A well-defined transition temperature can be established using a variety of observables that are moments of the order parameter, which for pure ${\rm SU}(3)$ theory is the Polyakov loop $P$. In the analysis of this section, we extract the transition temperature using two of these observables: the susceptibility of the Polyakov loop, \begin{equation}
\chi = N_s^3 \left(\langle \vert P \vert^2 \rangle - \langle \vert P \vert \rangle^2\right), \label{eq:susceptibility definition}
\end{equation} and the third-order Binder cumulant of the Polyakov loop, \begin{equation}
b_3 = \frac{\langle\vert P \vert^3\rangle - 3\langle\vert P \vert\rangle\langle\vert P \vert^2\rangle + 2\langle\vert P \vert\rangle^3}{\left(\langle \vert P \vert^2 \rangle - \langle \vert P \vert \rangle^2\right)^{3/2}}. \label{eq:b3 definition}
\end{equation} At finite volume, the susceptibility as a function of the coupling $\beta$ shows a peak around the transition temperature. The value of the coupling at which the peak occurs, which we call $\beta_\chi$, approaches a value $\beta_c$ in the infinite volume limit as the height of peak diverges linearly with the volume and the width of the peak goes to zero linearly with the inverse volume; this value corresponds to the transition temperature. $b_3$ is used in an analogous fashion: as an odd central moment of the order parameter, instead of a peak, $b_3$ has a zero-crossing at a particular coupling value which we call $\beta_b$. The slope of $b_3$ around the zero-crossing increases linearly with the volume. In the infinite volume limit, the zero-crossing of $b_3$ becomes a jump-discontinuity at the phase transition, and $\beta_b$ approaches the same value of the critical couplings as does $\beta_\chi$. 

If one wants to compare results at different lattice spacings, the susceptibility defined in Eq. (\ref{eq:susceptibility definition}) should be renormalized when one extracts $\beta_\chi$ \cite{Aoki:2006br}.  The zero-crossing of $b_3$ is clearly unaffected by scaling, so $b_3$ does not need to be renormalized when extracting $\beta_b$. For every lattice, we renormalize the bare susceptibility as
\begin{equation}
\chi_R(\beta,N_s, N_t) = Z^{N_t}(\beta)\, \chi(\beta,N_s,N_t), \label{eq:renormalization}
\end{equation}
where $N_t$ is the temporal extension of that lattice, with the renormalization condition that
\begin{equation}
\chi_R(\beta_b,4N_t,N_t)= Z^{N_t}(\beta_b)\, \chi(\beta_b,4N_t,N_t)=1, \label{eq:renormalization condition}
\end{equation}
i.e. we fix $\chi_R(\beta_b)=1$ for all lattices with aspect ratio $LT = N_s / N_t = 4$. $\beta_b$ is used in the renormalization condition because, as just mentioned, it is a scale-invariant quantity. Because the normalization function $Z(\beta)$ is itself $\beta$-dependent, the location of the peak of $\chi_R(\beta)$ differs slightly from that of $\chi(\beta)$; hence, $\beta_\chi$ is dependent on the renormalization.
Obviously, it is not strictly necessary to renormalize the susceptibility in order to extract $\beta_c$. As the width of the bare peak decreases linearly with the inverse volume, $Z(\beta)$ becomes approximately constant across that width, and so the difference between $\beta_\chi$ of $\chi(\beta)$ and $\beta_\chi$ of $\chi_R(\beta)$ should go to zero in the infinite volume limit. Thus, the value of $\beta_c$ extracted from $\chi(\beta)$ should agree with the value of $\beta_c$ extracted from $\chi_R(\beta)$. However, that our results may be formally correct and to avoid any potential ambiguities, we use the renormalized susceptibility in this analysis.

We extract $\beta_\chi$ and $\beta_b$ by fitting $\chi_R(\beta)$ and $b_3(\beta)$ in the vicinity of the peak and of the zero-crossing respectively. We employ basic polynomial fits of degrees 3, 4, 5, and 6. As a consequence of the stream-swapping in the tempering algorithm, correlations exist in the Polyakov loop data computed at different values of $\beta$ for a given lattice (see Sec. \ref{sec:tempering} for details); thus, correlations also exist among each moment of the Polyakov loop computed at different $\beta$ for each lattice. Therefore, we use the well-established practice of fitting with correlations using a subspace of the covariance matrix. The essential details are as follows. When fitting either $\chi_R(\beta)$ or $b_3(\beta)$ for a particular lattice, the covariance matrix of the fitted quantity is computed and diagonalized. In the diagonalized matrix, the smallest eigenvalues are averaged over and replaced with that average. After the averaging, the covariance matrix is converted back to its original basis. The cutoff for what counts as a small eigenvalue is freely chosen. We consider three choices for the cutoff: 10\%, 5\%, and 1\% of the largest eigenvalue. For completeness and comparison, we also consider an uncorrelated fit. Ultimately, whether one uses a correlated fit or no correlation at all, we find that it makes very little difference on the final result for the transition temperature (see the error budget).
In order to be able to interpret the $\chi^2$ of the fit to $\chi_R(\beta)$ or $b_3(\beta)$, one can use a correlated fit. The final result for the critical couplings
including combined systematic and statistical errors are included in App.~\ref{app:betacrittable}.

To extrapolate $\beta_\chi$ and $\beta_b$ to the continuum and infinite volume limits to get $\beta_c$, we use the following set of lattice parameters:

\begin{table}[H]
\centering
\centering
\begin{tabular}{l|l}
\toprule
$N_t$  &  $N_s$ \\
\hline 
5      &  15, 20, 22*, 23*, 25, 30, 40 \\
6      &  18, 21, 24, 27, 30, 36, 48 \\
7      &  28 \\
8      &  24, 28, 32, 36, 40, 48, 64 \\
10     &  30, 35, 40, 45, 50, 60, 80 \\
12     &  48 \\
\toprule
\end{tabular}
\label{tbl:table of lattice parameters}
\end{table} \noindent The volumes are chosen such that there are lattices with aspect ratios $LT = 3$, 4, 4.5, 5, 6, and 8, for $N_t = 5$, 6, 8, and 10 (to obtain data at $LT = 4.5$ for $N_t = 5$, we find $\beta_\chi$ and $\beta_b$ for the lattices $22^3{\times}5$ and $23^3{\times}5$ and then interpolate linearly in $1/N_s^3$ to get $\beta_\chi$ and $\beta_b$ for a hypothetical ``$22.5^3{\times}5$'' lattice), and that there are lattices with $LT = 3.5$ for $N_t = 6$, 8, and 10. Two additional lattices of $LT = 4$ are also chosen for $N_t = 7$ and 12.

Now, the value of the coupling $\beta$ is specific to the choice of action on the lattice; hence, while $\beta$ is useful for comparing results computed using the same action, it must be translated into a physically meaningful quantity, the lattice temperature $T$, to yield results that are independent of the choice of action. This requires the use of a scale setting. We use the $w_0$ scale based on the Wilson flow, implicitly defined by \begin{equation}
t \frac{d}{dt}\left[ t^2 E(t) \right]_{t = w_0^2} = 0.3 , \label{eq:w0 scale definition}
\end{equation} where $E(t)$ is the expectation value of the gauge action of lattice configurations evolved via the Wilson flow \cite{Borsanyi:2012zs, Luscher:2010iy}. We compute the $w_0$ scale in lattice units, $w_0/a$, for many values of $\beta$ for two different discretisations of the flow
(WSC and SSC in the notation of \cite{Fodor:2014cpa}).
This gives us another systematic choice as to which version of the scale setting to use.
We then interpolate these results to get $w_0/a(\beta)$ by fitting with a polynomial of order 6 and 7 in the $\beta$ range [4.0,4.95].  It is critical to ensure that finite-volume effects on the $w_0$ scale remain small, as these effects increases with the flow time \cite{Borsanyi:2012zs}. In Fig. \ref{fig:w0/a volume dependence}, we present a volume study of the $w_0$ scale (WSC) computed in this analysis. Fluctuations in $w_0/a$ are comparable in size to the statistical error at each volume; consequently, no significant volume dependence can be seen.

Once $w_0/a(\beta)$ is found for a particular $\beta$ on a lattice with a temporal extension of $N_t$, it can be converted into the lattice temperature $T = (N_t a)^{-1}$ by dividing by $N_t$: \begin{equation}
\frac{w_0}{aN_t} = w_0 T. \label{eq:w0/a to temperature}
\end{equation} In this way, $\beta_\chi$ and $\beta_b$ are converted into the dimensionless quantity $w_0 T_c$ for each lattice.

To compute the continuum value of the transition temperature $w_0 T_c$ at infinite volume,
we first find the continuum value of $w_0 T_c$ at finite volume and fixed lattice aspect ratio $LT = N_s / N_t$ and then extrapolate the results of the continuum theory at different $LT$ to the limit where $(LT)^3 \rightarrow \infty$. That a continuum limit for $w_0 T_c$ exists for pure ${\rm SU}(3)$ at finite volume is not immediately obvious; however, as shown in Fig. \ref{fig:w0*T continuum limit}, a well-defined continuum value of $w_0 T_c$ can in fact be established at fixed aspect ratio $LT$ using both definitions for $\beta_c$.
We use large lattices with aspect ratios $LT = 4$, 4.5, 5, 6, and 8 for four different temporal extensions: $N_t = 5$, 6, 8, and 10. To check the consistency of the continuum extrapolation at other temporal extensions, we also include two lattices with $N_t = 7$ and 12 in the $LT = 4$ extrapolation. In the left panel of Fig. \ref{fig:w0*T continuum limit}, one can see that $w_0 T_c$ goes roughly linearly with  $N_t^{-2}$ for each $LT$. We perform two types of fit to extrapolate to the continuum: a linear fit to all lattices with $N_t \geq 6$, and a quadratic fit to all lattices with $N_t \geq 5$.  The fits are shown in Fig. \ref{fig:w0*T continuum limit}; the error bands give the combined statistical and systematic errors from all the similar fits performed in the analysis.

The continuum results from $\chi_R$ and $b_3$ are shown in Fig. \ref{fig:w0*T infinite volume limit} as functions of $(LT)^{-3}$, which behaves as the inverse volume. A quadratic fit to $LT \geq 4$ is used to extrapolate to the infinite volume limit. This fit is shown in Fig. \ref{fig:w0*T infinite volume limit} for both $\chi_R$ and $b_3$. The error bands give the combined statistical and systematic errors from all the similar fits that were performed. We find that the infinite volume value of the transition temperature $w_0 T_c$ computed using $\chi_R$ agrees with the value computed using $b_3$, as expected.

Following the analysis method introduced in \cite{Borsanyi:2020mff}
to estimate the statistical and systematic uncertainties of the results,
in total we have performed 256 different analyses, these are
characterized by the choice of the moment of the Polyakov loop ($\chi_R$ or $b_3$), the degree of the fit to the moment (3, 4, 5, or 6), whether and how the fit is correlated (uncorrelated or correlated with an eigenvalue cutoff of 10\%, 5\%, 1\%), the choice of the $w_0$ scale calculation (WSC, SSC), the degree of fit to the $w_0$ scale data (6 or 7), and the degree and range of the continuum extrapolation (linear or quadratic). The cumulative distribution function (CDF) of the infinite volume results of these analyses is shown in Fig. \ref{fig:w0*T CDF} in red, where the central value and statistical error of each result are given. The statistical and systematic uncertanities are synthesized into a smooth CDF defined by the equation \begin{equation}
F(w_0 T_c) = \frac{1}{2} + \frac{1}{2\cdot 256} \sum_{i=1}^{256} w_i \, {\rm erf}\left( \frac{w_0 T_c - \mu_i}{\sqrt{2} \sigma_i} \right) \label{eq:CDF definition}
\end{equation} where $\mu_i$ and $\sigma_i$ are the central value and statistical uncertainty respectively of the $i$th analysis, and $w_i$ is the weight of the CDF of the $i$th result. (This equation then assumes that the statistical result of each analysis is well-described by a normal distribution with a mean of $\mu_i$ and standard deviation of $\sigma_i$.) We have no prior assumptions as to the relative statistical significance of any of the analyses, and so we take an agnostic position by weighting all analyses equally, i.e. by setting $w_i = 1$. The result is shown in blue in Fig. \ref{fig:w0*T CDF}. The final result is then the median value of the smooth CDF, implicitly defined by $F(w_0 T_c) = 0.5$, and the central 68\% width is taken as twice the total error. This yields a final value of $w_0 T_c = 0.25384(23)$, which is shown in green in Fig \ref{fig:w0*T infinite volume limit}, with an error budget of 
\begin{tabular}{l|l|l}
\hline\hline
median                         & \multicolumn{2}{c}{ 0.25384} \\
statistical error              &  0.00011 & 0.043 \% \\
full systematic error          &  0.00021 & 0.082 \% \\
\hline
 Observable ($b_3$ , $\chi_R$)            &  $ 1.1 \cdot 10^{-5}$ & 0.0042 \% \\
 Fit order (3, 4, 5, 6)         &  $ 2.1 \cdot 10^{-5}$ &   0.0085 \% \\
 Fit type (corr, uncorr)  & $ 1.2 \cdot 10^{-5} $&   0.0047 \% \\
 Scale setting (WSC, SSC) & $  1.9 \cdot 10^{-5}$ &   0.0075 \% \\
 Scale fit order (6, 7)    & $ 2.4 \cdot 10^{-6}$ &   0.0010 \% \\
 Continuum limit range &  $ 1.2 \cdot 10^{-4}$ &   0.0487 \% \\
\hline\hline
\end{tabular}


\section{\label{sec:firstorder}Evidence for a first order transition} 

In this section we use our simulations to demonstrate that the
thermodynamic transition of the SU(3) Yang-Mills theory is
first order, as anticipated. We can perform this in two ways.

First we study the finite volume scaling of the susceptibility
(scaled variance) of the order parameter. As we did in the case
of the transition temperature study, we again use the susceptibility 
($\chi_R$) of the renormalized Polyakov loop. If the transition is of first
order, the leading volume dependence must be $\chi_R^{-1} \sim V^{-1}$.
Such a study would not be new and also would not give a true evidence,
if we used data from a fixed lattice resolution. Here we first calculate
the continuum limit of the peak height for various volumes, and
then demonstrate the expected volume scaling. Moreover, we show the 
continuum extrapolation of $\chi_R$ as a function of $w_0 T$ around the 
critical value $w_0 T_c$, and display the expected volume dependence.

Perhaps the most convincing argument for the first order nature
of a transition is the existence of a latent heat in the
thermodynamic and continuum limit. Here we are
building on the work of the WHOT collaboration
\cite{Shirogane:2016zbf,Shirogane:2020muc}. Aided by the
tempering algorithm and the improved action we arrive at
a combined continuum and infinite volume limit, 29
standard deviations away from zero.


\subsection{Finite volume scaling of the Polyakov loop susceptibility} 

The obtain the finite volume scaling of a susceptibility, it has to be
continuum extrapolated for several fixed physical volumes.
This raises the question of renormalization, a topic that one could ignore
as long as only one lattice resolution is relied upon. As discussed
before, $\chi_R(L,T) = Z^{N_t}(\beta)\chi(\beta,N_x,N_t)$ with
$L=a(\beta) N_x$ and $T^{-1}=a(\beta) N_t$.
The renormalization condition for $Z$ is defined at a fixed volume, for us the choice is $LT=4$ (meaning $N_x=4N_t$).
Scanning through the resolutions $N_t=4,5,6,7,8,10,12$ and setting the
renormalization condition $\chi_R(L=4 T_c^{-1},T=T_c) \equiv 1$ one
can easily obtain $Z(\beta)$ for the whole range of interest.
Clearly, in this setting we lose any information about an overall constant
in $\chi_R$, but being this a volume independent factor, the actual
finite size scaling is left intact.

\begin{figure*}
\includegraphics[width=0.99\textwidth]{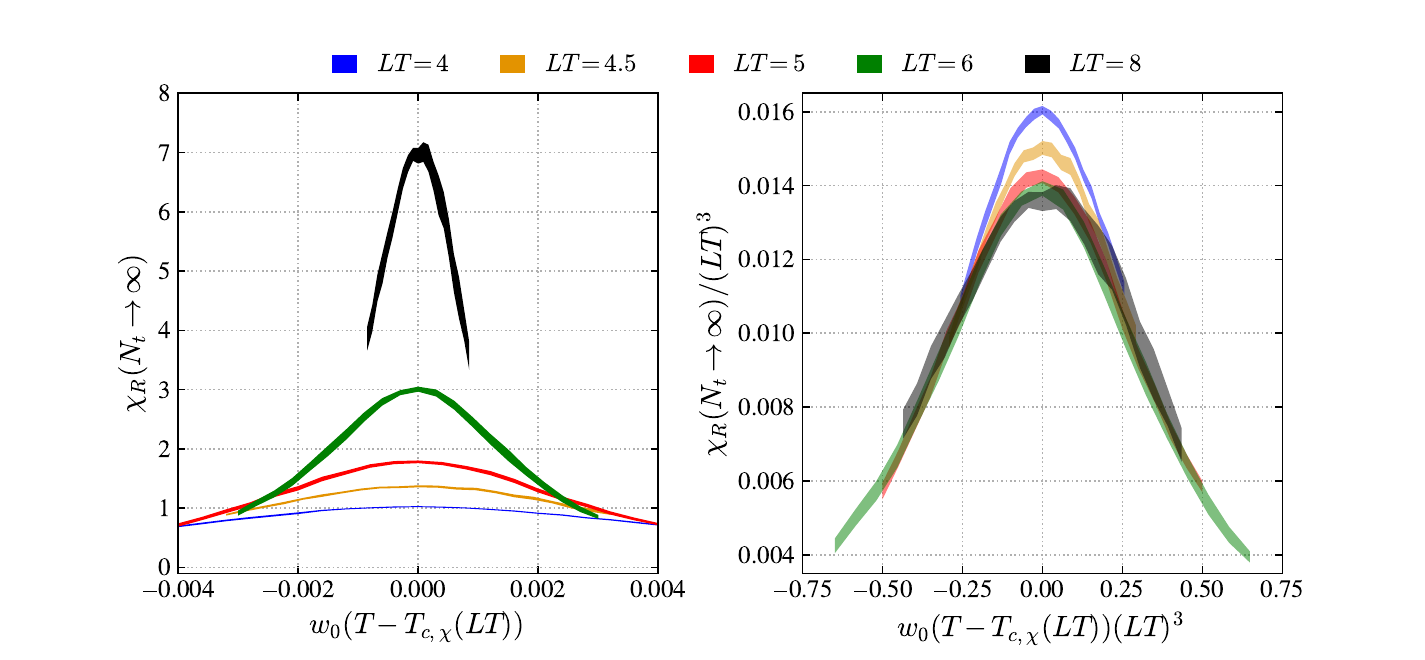}
\caption{\label{fig:chiR_contlim}
(Left) The continuum extrapolated renormalized Polyakov loop susceptibility at different
volumes, centered around the volume-dependent peak position $w_0T_{c,\chi}(LT)$. The 
bands include both statistic and systematic uncertainties.  (Right) Same curves as in the 
left panel, but scaled with appropriate powers of the volume, in order to highlight the 
linear volume dependence typical of a first order transition. 
}

\end{figure*}

\begin{figure}
\centering
\includegraphics[width=0.88\linewidth]{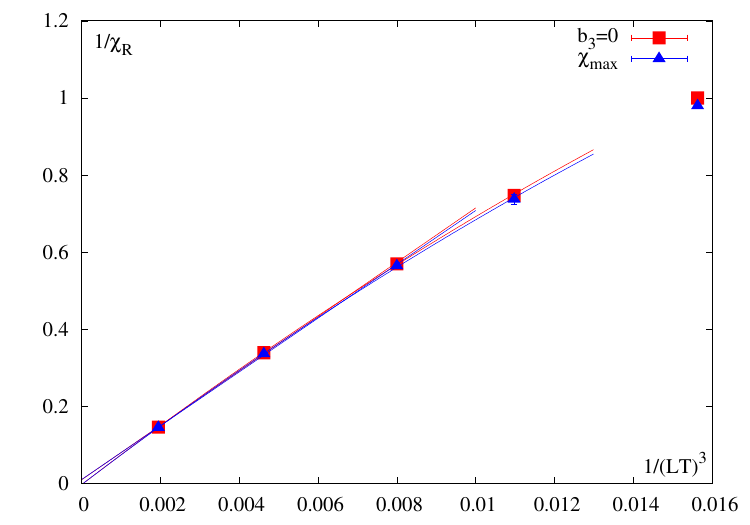}
\caption{\label{fig:chifss}
Infinite volume extrapolation of the inverse Polyakov loop susceptibility.
The results are renormalized and continuum extrapolated, 
systematic errors are included.
The renormalization condition was $ \chi_R(L=4T_c^{-1},T_c)=1$.
We use red symbols if we defined
$T_c$ though the zero of the third Binder cumulant, blue symbols when we defined
$T_c$ as the peak of the susceptibility.
Depending on the details of the infinite volume extrapolation we get a slightly
positive or negative result. 
All in all we get the result
$ \chi_R^{-1}(V=\infty)= 0.0023 (58)_{\rm stat}(65)_{\rm sys} $ 
The linear scaling sets in for $LT\ge5$.
The linear convergence of $ \chi_R^{-1} $ to zero indicates a first order transition.
}
\end{figure}

We select fixed physical volumes ($LT=4.5, 5, 6, 8$) and interpolate $Z(\beta)$ to
the actual volume dependent $\beta_c$ of each simulation with $N_t=5, 6, 8$~and~10.
The renormalized result $\chi_R(L,T)=Z^{N_t}(\beta) \chi(\beta,N_x,N_t)$
is simply continuum extrapolated as~$N_t^2$.
Such continuum extrapolation is shown in Fig.~\ref{fig:chiR_contlim} for the 
$\chi_R(w_0T)$ curves at fixed volume, and in Fig.~\ref{fig:chifss} for the inverse peak 
height.
In Fig.~\ref{fig:chiR_contlim} (left), each curve is continuum extrapolated at the indicated
fixed volume, and the bands include both statistic and systematic uncertainties. Note 
that, had we subtracted the infinite volume limit value $\chi_R(w_0T_c)$ instead of the 
volume-dependent peak values $\chi_R(w_0T_{c,\chi}(LT))$, the curves would not have 
have been exactly centered around zero, but a trend would have appeared where larger 
volumes would peak closer to zero. However, as can be seen in the right panel of 
Fig.~\ref{fig:w0*T infinite volume limit}, the difference would be negligible, since the 
continuum extrapolated peak position shows a very mild volume dependence.  The right
panel shows the same curves, but scaled with the volume in order to highlight the 
volume dependence. We can see that the bands corresponding to $LT=5,6,8$ are overall
almost indistinguishable, clearly showing that they fall in the linear volume scaling 
regime typical of a first order transition.

In Fig.~\ref{fig:chifss} we actually show two extrapolations, one for each definition
of $T_c$, using the zero of the 3rd Binder cumulant or the susceptibility peak in red and blue, 
respectively. The extrapolations do not differ at all depending on this ambiguity, the
main systematic uncertainty comes from the
possibility to include a inverse quadratic volume term in the infinite volume
extrapolations. We find that the
inverse susceptibility is extrapolated to be vanishing, (or, actually, strongly constrained: $ \chi_R^{-1}(V=\infty)= 0.0023 (58)_{\rm stat}(65)_{\rm sys} $  )
in the infinite volume limit. (A crossover transition is signaled by a positive
value of the infinite volume extrapolated inverse susceptibility).

\subsection{\label{sec:latent}Calculation of the latent heat} 

\begin{figure}
\centering
\includegraphics[width=\linewidth]{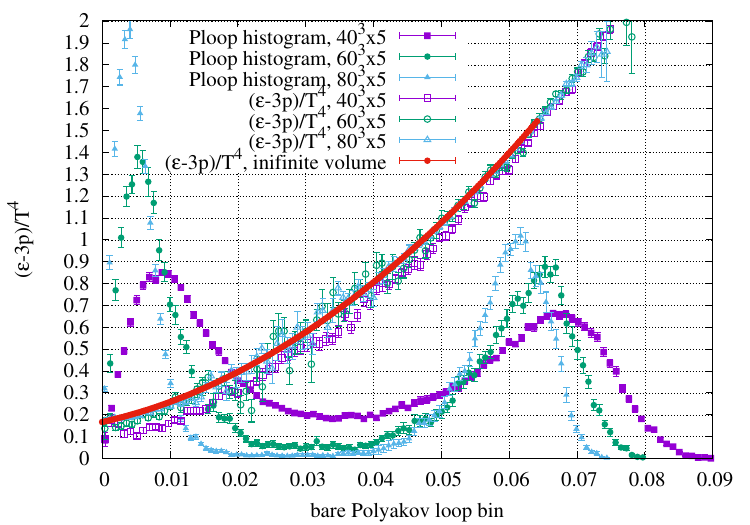}
\caption{\label{fig:latent5}
We show the Polyakov loop histograms for three volumes ($N_t=5$) along
with the trace anomaly expectation values for each Polyakov loop bin.
The double peak structure and the sharpening of the peaks
with increasing volume match our expectations for a first order
transition. The trace anomaly shows
no discontinuity in this representation, the infinite volume limit
can be found using a polynomial model.
}
\end{figure}

A non-vanishing latent heat identifies a transition to be first order
beyond doubt. Its calculation, however, is much less trivial than
observing a transition peak in a susceptibility. Most conveniently,
the latent heat can be understood as the discontinuity of the
trace anomaly, $(\epsilon -3p)/T^4$, as the transition is crossed.
This assumes a continuous pressure, which is assumed in all transition
scenarios. The difficulty in quantifying this gap arises from two sources:

a) To calculate a discontinuity one must get very close to $T_c$
from both sides. The trace anomaly happens to have a pronounced
temperature dependence just above and just below $T_c$. 

b) The trace anomaly has a quartic divergence, thus
brute force statistics must scale with $\sim N_t^8$ to compensate.
Even in quenched QCD, this can quickly become the driving factor 
in the costs for $N_t\ge 8$.

When the continuum limit is desired, point b) determines the amount of statistics needed, 
but point a) must be addressed in all cases.  Just ``reading off the jump'' could give any
value between zero and 2 for $\Delta (\epsilon -3p)/T_c$ as shown in Fig.~1
of our earlier study \cite{Borsanyi:2012ve}.
One could define the latent heat as the surface under the peak of the energy
susceptibility, however, the signal-to-noise ratio is diminishing
in the infinite volume limit.

The most successful method to define the latent heat, the one we also use here,
is to simulate right at $\beta_c$ and classify the lattice configurations
into the cold and hot phases (see Refs. \cite{Shirogane:2016zbf,Shirogane:2020muc}).
The trace anomaly can then be evaluated for both the hot and cold halves of the ensemble,
the difference gives the latent heat up to a renormalization factor.
The absolute value of the Polyakov loop shows a clear double peaked histogram.
Selecting the Polyakov loop magnitude at the minimum of this histogram
between the two peaks as the cut value, we introduce a phase categorization
which has a systematic error decreasing exponentially
with growing volume.
This criterion works well on smeared (flowed) configurations as well \cite{Shirogane:2020muc}.

We illustrate this approach on our $N_t=5$ ensembles at $\beta_c$. We reweighted our
closest $\beta$-ensemble to the point where $b_3(\beta)=0$ for the particular
simulation volume. The bare Polyakov loop histograms are shown in
Fig.~\ref{fig:latent5} for three of our simulation volumes. We show the standard
jackknife errors on the histograms. For this lattice spacing
we could reach an aspect ratio of $LT=16$, where the separation of the two
peaks are very clear, indeed. Yet even near the minimum we have enough hits 
in the Polyakov loop bins to quantify the trace anomaly bin by bin.

The trace anomaly shows no discontinuity between the phases, its Polyakov loop
dependence can be modeled by a polynomial. In Fig.~\ref{fig:latent5} we
fit a 3rd order polynomial with finite volume
corrections proportional to $1/L^3$ to obtain the infinite volume extrapolation (red curve).
This picture connects the latent heat with the hot-phase value of Polyakov loop.
The cold phase peak moves to zero as $1/L^3$, hence, the $y-$intercept of the
red curve points to the trace anomaly at $T_c-\epsilon$. Similarly, the
non-trivial position of second histogram peak in the thermodynamic limit is
translated by the same curve to the trace anomaly at $T_c+\epsilon$.

As observed in Fig.~\ref{fig:latent5}, one may define the latent heat at a fixed
ensemble, by translating the peak positions into a pair of
trace anomaly values through the red curve (the infinite volume extrapolation
of the trace anomaly as function of the magnitude of the Polyakov loop), and than taking
the difference.
While this is clearly possible, it is to be seen in practice, if the volume
scaling turns linear in $1/L^3$ for practical lattice sizes. Alas, this is not the case.
Both the peak positions and the trace anomaly they relate to, are heavily non-linear
in the range $4\le LT\le 8$, where most of our statistics is collected.
However, integrating the trace anomaly curve with the histogram peak's weight
for both peaks gives a linear result. This is then equivalent to the approach
in Ref.~\cite{Shirogane:2016zbf} where the trace anomalies were simply averaged
for both cold and hot sub-ensembles.
In summary, we first define a cut value in the modulus of the Polyakov loop for each lattice size individually
as the minimum position of the discussed histogram. These bins play no role in the second step, where we calculate the trace anomaly for the configurations above and below this cut, separately.
The trace anomaly difference between these two sub-ensembles gives the latent heat normalized to $T^4$.

Let us now come to point b). Ref.~\cite{Shirogane:2020muc} introduces a truly innovative
way to separate the UV noise of the physics result in the trace anomaly. The results
are calculated at small gradient flow time, where the square root of the flow time
replaces the lattice spacing in its role as a cut-off scale. The presence of a
small, physical flow time eliminates the quartic divergence and the continuum limit can be taken.
Using larger flow time allows the usage of coarser lattices.
This then needs to be related
to the real theory at zero flow time using small flow time expansion \cite{Suzuki:2013gza},
which gets harder as the selected flow time increases.

In this study, we analyzed the configurations at zero flow time directly. To mitigate
the quartic divergence, we used the tree-level Symanzik improved action. This choice does
not improve the quartic scaling of the simulation noise, yet it endorses coarser lattices
within the linear range continuum scaling. We use lattices up to $N_t=10$ instead of 
the $N_t=16$ of Ref.~\cite{Shirogane:2020muc}, this immediately allows for simulating
a factor $43\approx(16/10)^8$ less statistics due to
the quartic divergence of the trace anomaly.
(This factor must be considered together
with the fact that the improved action comes at a 2.1 times higher cost\footnote{
The 2.1 factor was measured without communication on an AMD EPYC node using the AVX-2 instructions. 
The operation count needed to calculate the staple 
is 7 times higher with the $2\times 1 $ term in the gauge action
than without.
}.)

The trace anomaly in Fig.~\ref{fig:latent5} was renormalized using the standard
scheme of Ref.~\cite{Boyd:1996bx}
\begin{equation}
\frac{\epsilon-3p}{T^4} = N_t^4 a \frac{d\beta}{da} \left(
\left\langle S_g\right\rangle_{N_s^3\times N_t}-
\left\langle S_g\right\rangle_{\mathrm{infinite}~N_t}
\right)\,,
\end{equation}
where $S_g$ is the gauge action without the factor of the inverse coupling $\beta$.
The logarithmic derivative of the $\beta$ -- lattice spacing ($a$) relation was
obtained using our $w_0$ scan, the same that already used as high precision scale
setting in Section~\ref{sec:tc}. The same runs were used to calculate the zero
temperature gauge action $\left\langle S_g\right\rangle_{\mathrm{infinite}~N_t}$, that
is not necessary for latent heat itself, only if one is interested in trace anomalies
at $T_c\pm \epsilon$.

\begin{figure}
\centering
\includegraphics[width=0.88\linewidth]{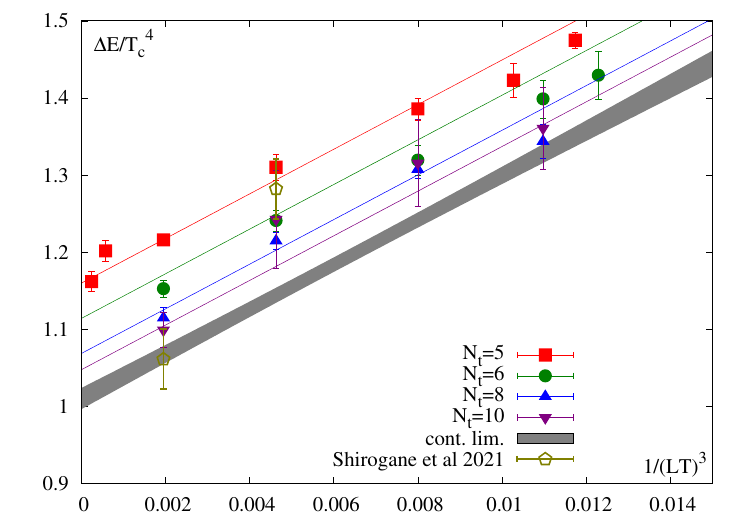}
\caption{\label{fig:latent_contlim}
Combined and continuum limit of the trace anomaly difference 
between the hot and cold phases. The extrapolation is linear both
in $1/N_t^2$ and $1/L^3$, in this example we exclude all ensembles with $N_t<6$ or $LT<5$.
The continuum extrapolated but finite volume results in Ref.~\cite{Shirogane:2020muc}
are shown for comparison.
}
\end{figure}

We calculated the trace anomaly differences for the ensembles that we generated. We find
that a combined linear continuum and infinite volume fit is possible and thus we
get the result in Fig.~\ref{fig:latent_contlim}.

For the final results we consider the following sources of systematic uncertainties.
The continuum limit is calculated both using and not using $N_t=5$, including
or not including the aspect ratio $LT=4.5$ into the infinite volume extrapolation.
We use various number of Polyakov bins (100, 200 or 400) and two different
log-polynomial models for fitting the negative logarithm of the histogram so that
cut value separating the phases can be calculated.
We have performed two separate analyses: first using reweighting
to $\beta_c$ from the closest available $\beta$ ensemble, second, using
the multihistogram method to calculate averages at $\beta_c$.
In the end we obtain

\begin{equation}
\Delta \left[\frac{\epsilon -3p}{T^4}\right] = 1.025(21)_{\rm (stat)} (27)_{\rm (sys)}
\end{equation}

with an error budget of

\begin{center}
\begin{tabular}{l|l|l}
\hline\hline
median                         & \multicolumn{2}{c}{  1.0249} \\
statistical error              &    0.021 &  2.1 \% \\
full systematic error          &    0.027 &  2.7 \% \\
\hline
 Histogram fitting               &   0.0030 &   0.29 \% \\ 
 Histogram binning              &   0.0002 &   0.02 \% \\
 $ \beta_c$ definition          &   0.0135 &   1.32 \% \\
 $w_0$ interpolation            &   0.0007 &   0.07 \% \\
 $LT$ range               &   0.0121 &   1.18\% \\
 $N_t$ range                &   0.0157 &   1.53\% \\
 Analysis method                &   0.0019 &   0.18\% \\
\hline\hline
\end{tabular}
\end{center}


\section{\label{sec:conclusions}Conclusion} 

In this paper we have studied the phase transition in an SU(3) pure gauge theory. First we have investigated the efficiency of parallel tempering
simulations to measure properties of the system near the phase
transition point. The parallel tempering allows swap updates
between simultaneous calculations at different temperatures.
One observes a sizable decrease of the autocorrelation
times compared to simulations at a fixed temperature.
As observed the efficiency of parallel tempering
is superior to even brute force simulations using multihistogram analysis
to combine the data of all the ensembles.

Next we have determined the critical temperature of the theory.
We used two definitions to find the transition point:
the peak of the Polyakov loop susceptibility and the zero of the
3rd Binder cumulant of the Polyakov loop.
The ensembles in this study consisted of lattices with temporal lattice size $ N_t=5,6,8,10 $ and 12 with aspect ratios $ LT = 4, 4.5,6,8,10,12$, using the Symanzik gauge action.
First, we carried out a continuum extrapolations for fixed aspect ratios,
giving the continuum extrapolated critical temperatures for finite physical
volumes. Second we calculated the infinite volume extrapolated
critical temperatures, and  and observed the coincidence of the
two definitions extrapolated to the thermodynamic limit in the
continuum, as expected.
The systematic errors of the calculation
were thoroughly investigated by exploring many possible fitting
and extrapolation procedures.
Our final result for the critical temperature
with combined statistical and systematic error is:  $w_0 T_c = 0.25384(23)$.

Finally we have investigated the properties of the
phase transition. We have confirmed the first order nature
of the transition by first investigating the finite volume scaling
of the peak of the Polyakov loop susceptibility. For this we have calculated 
the continuum limit of the renormalized Polyakov loop susceptibility
for different physical system sizes.
As one observes both the height and the width of the susceptibility follow the 
appropriate scaling in the infinite volume limit, confirming that
the phase transition is first order.
Second, we have calculated the latent heat of the transition, with the
help of the discontinuity of the trace anomaly. The discontinuity was measured
for each ensemble by averaging the trace anomaly at the critical coupling
for ``hot'' and ``cold'' configurations, where the classification
is based on the value of the Polyakov loop.
Finally we carried out a combined infinite volume and continuum extrapolation,
yielding the value $ \Delta \epsilon / T_c^4 = 1.025(21)(27) $.

The tempering algorithm is not limited to quenched QCD. The same first
order transition and the corresponding critical end-point can be addressed
using tempering updates in gauge coupling in the heavy quark region
\cite{Kara:2021btt}. Introducing pseudofermions the tempering updates
can be extended to mass, imaginary chemical potential or other parameters of
the fermionic action. This algorithm will probably play an important
role in the exploration of the QCD phase diagram.

\textbf{Acknowledgments}
The project received support from the BMBF Grant No. 05P21PXFCA.  The
authors gratefully acknowledge the Gauss Centre for Supercomputing e.V.
(www.gauss-centre.eu) for funding this project by providing computing time on
the GCS Supercomputer HAWK at HLRS, Stuttgart.  Part of the computation was
performed on the QPACE3 funded by the DFG and hosted by JSC and on the
cluster at the University of Graz. 
S.B. thanks Julius Kuti for discussions on the topic.


\appendix

\section{\label{app:corrmulti}Correlated Multihistogram}

Here we adopt the well known reweighting method by Ferrenberg and Swendsen \cite{Ferrenberg:1989ui} and extend it to correlated data.\\
The weighted average $a$ of data sets $x_i$, $i=1,2,...,m$ can be calculated by minimizing
\begin{equation}
\chi^2 = \sum_{i,j}^m (x_i -a) \left( C^{-1} \right)_{i,j} (x_j -a),
\end{equation} 
with $C^{-1}$ being the inverse covariance matrix \cite{Schmelling:1994pz}. The minimum is found for
\begin{equation}
a=\frac{   \sum_{i,j=1}^{m} \left( C^{-1} \right)_{i,j} x_j   } { \sum_{i,j=1}^{m} \left( C^{-1} \right)_{i,j} },
\label{eq:weight_sum}
\end{equation}
which has minimal fluctuations compared to other possible linear combinations.\\
Interpolating observables such as the Polyakov loop to a target coupling $\beta$ demands a precise knowledge of the partition function $Z(\beta)$ and thus of the density of states $W$.
\begin{equation}
Z(\beta)=\sum_E W(E) e^{-\beta E}
\label{eq:Z}
\end{equation}
Since $W(E)$ does not depend on $\beta$ it can be estimated by the densities of states of the individual simulations $W_i(E)$ according to eq.~(\ref{eq:weight_sum}). Thus $i$ labels the simulation at $\beta_i$ which is correlated with the other ones due to the $\beta$ tempering algorithm. The histograms of the energy $N_i(E)$  can be related to the density of states by
\begin{equation}
W_i(E) \frac{e^{-\beta_i E}}{ Z(\beta_i) } = \frac{N_i(E)}{n_i},
\end{equation} 
where $n_i$ is the total amount of entries of the histogram. In the same manner the exact density of states $W(E)$ is estimated by multiple simulations performed at the same coupling
\begin{equation}
W(E)=\frac{ \overline{N_i}(E) Z(\beta_i)  }{n_i e^{-\beta_i E}}.
\end{equation}
It is important to note that $\overline{N_i}$ is the averaged histogram of these simulations, thus its error is $\sigma_i=\sqrt{g_i \overline{N_i}}$. The factor $g_i$ takes the correlation time $\tau_i$ of the ensemble into account and reads $g_i=1+2\tau_i$. 
The covariance for a certain energy $E$ between the densities of states of individual simulations $W_i$ and $W_j$ can be calculated as
\begin{equation}
\mathrm{cov}(W_i,W_j)=W^2  \frac{\mathrm{cov}(N_i,N_j)} {\overline{N_i} \; \overline{N_j}} = W^2 \frac{\mathrm{corr}(N_i,N_j) g_i g_j} {\sigma_i \sigma_j}.
\end{equation}
Here $\mathrm{cov}(N_i,N_j)$ and $\mathrm{corr}(N_i,N_j)$ stand for components of the covariance and correlation matrix respectively which are estimated according to the standard Jackknife resampling method.\\
With eq.~(\ref{eq:weight_sum}) the estimator of the exact density of states $W(E)$ can be written as
\begin{align}
\nonumber & W(E)=\frac{   \sum_{i,j=1}^{m}  \sigma_i \sigma_j (g_i g_j)^{-1}  \;  \mathrm{corr}^{-1}(N_i,N_j) \; W_j   } { \sum_{i,j=1}^{m}  \sigma_i \sigma_j  (g_i g_j)^{-1} \; \mathrm{corr}^{-1}(N_i,N_j)  }\\
&=\frac{\sum_{i,j=1}^{m}  (g_i g_j)^{-1/2}   \sqrt{\frac{n_i Z_j}{n_j Z_i}} e^{ E(\beta_j - \beta_i)/2} \;  N_j(E)  \; \mathrm{corr}^{-1}(N_i,N_j) 
}{\sum_{i,j=1}^{m} (g_i g_j)^{-1/2} \sqrt{\frac{n_i n_j}{Z_j Z_i}} e^{ -E(\beta_i + \beta_j)/2  } \; \mathrm{corr}^{-1}(N_i,N_j)  }.
\label{eq:deos_corr}
\end{align}
In some cases the correlation matrix is singular, caused by bins which only contain a few entries. Since these bins have a very small contribution to the end result of $W(E)$, we neglect the correlation of them with other bins.
In the case of uncorrelated simulations the correlation matrix is the unity matrix and we get back to the standard Ferrenberg-Swendsen equation for the density of states
\begin{equation}
W(E)=\frac{\sum_{i=1}^m g_i^{-1} N_i(E)} {\sum_{i=1}^m  g_i^{-1} \frac{n_i}{Z_i} e^{-\beta_i E} }.
\label{eq:standard_ferrenberg}
\end{equation}
\noindent
Inserting eq.~(\ref{eq:deos_corr}) in eq.~(\ref{eq:Z}) the partition functions $Z_i$ of the individual simulations can be calculated iteratively.
Assuming constant auto-correlation times $\tau_i$ the $g_i$ factors are the same for every simulation.
According to eqs.~(\ref{eq:deos_corr}) or (\ref{eq:standard_ferrenberg}) they represent a constant factor for all $Z_i$ which can be dropped.


\section{Critical couplings }
\label{app:betacrittable}

The critical couplings for the ensembles used in this study are listed
in Table~\ref{betacrittable}.
The quoted errors are combined systematic and statistical errors for the $\beta_c $ values
calculated using fitting as described in Sec.~\ref{sec:tc}.
Note that the critical couplings calculated
using the multihistogram analysis have no systematic error, as they are simply solutions of the eqs. $ b_3(\beta_b)= \partial_\beta \chi(\beta_\chi) =0 $).

\onecolumngrid

\begin{table}
\begin{tabular}{|c|c|c|c|c|c|}
\hline
Lattice & $\beta_c$ from $b_3$,fit  & $\beta_c$ from $b_3$, multihist & $\beta_c$ from $\chi^R $,fit  & $ \beta_c$ from $\chi$,fit & $\beta_c$ from $\chi$, multihist \\\\
\hline
$ 15 \times 5$ & 4.19667(15) & 4.19676(13) & 4.20246(73) & 4.20019(46) & 4.19996(16) \\
$ 20 \times 5$ & 4.19855(9) & 4.19858(8) & 4.20048(47) & 4.19987(14) & 4.19998(10) \\
$ 22 \times 5$ & 4.19917(5) & 4.19915(6) & 4.20050(28) & 4.20020(9) & 4.20020(6) \\
$ 23 \times 5$ & 4.19926(9) & 4.19930(9) & 4.20073(13) & 4.20040(12) & 4.20026(9) \\
$ 25 \times 5$ & 4.19995(11) & 4.19996(11) & 4.20084(21) & 4.20064(16) & 4.20069(11) \\
$ 30 \times 5$ & 4.20056(11) & 4.20052(10) & 4.20103(13) & 4.20095(13) & 4.20090(10) \\
$ 40 \times 5$ & 4.20112(3) & 4.20110(4) & 4.20126(6) & 4.20123(3) & 4.20123(4) \\
\hline
$ 18 \times 6$ & 4.30962(39) & 4.30986(41) & 4.31706(116) & 4.31412(97) & 4.31324(60) \\
$ 21 \times 6$ & 4.31081(33) & 4.31090(29) & 4.31470(72) & 4.31325(53) & 4.31319(33) \\
$ 24 \times 6$ & 4.31210(19) & 4.31190(12) & 4.31434(28) & 4.31355(28) & 4.31361(12) \\
$ 27 \times 6$ & 4.31233(9) & 4.31230(12) & 4.31389(16) & 4.31350(14) & 4.31347(12) \\
$ 30 \times 6$ & 4.31291(13) & 4.31290(15) & 4.31402(19) & 4.31376(17) & 4.31374(16) \\
$ 36 \times 6$ & 4.31379(11) & 4.31373(10) & 4.31427(17) & 4.31415(10) & 4.31417(10) \\
$ 48 \times 6$ & 4.31432(3) & 4.31432(5) & 4.31447(4) & 4.31445(3) & 4.31447(5) \\
\hline
$ 24 \times 8$ & 4.50511(26) & 4.50506(24) & 4.51309(77) & 4.50954(47) & 4.50936(38) \\
$ 28 \times 8$ & 4.50633(23) & 4.50639(20) & 4.51191(46) & 4.50980(43) & 4.50964(37) \\
$ 32 \times 8$ & 4.50756(14) & 4.50758(13) & 4.51060(34) & 4.50965(19) & 4.50964(12) \\
$ 36 \times 8$ & 4.50788(6) & 4.50795(7) & 4.50985(7) & 4.50934(9) & 4.50936(8) \\
$ 40 \times 8$ & 4.50846(9) & 4.50847(10) & 4.50985(16) & 4.50951(12) & 4.50952(11) \\
$ 48 \times 8$ & 4.50929(10) & 4.50930(9) & 4.50996(12) & 4.50982(11) & 4.50982(9) \\
$ 64 \times 8$ & 4.51028(7) & 4.51024(7) & 4.51048(8) & 4.51046(8) & 4.51041(7) \\
\hline
$ 30 \times 10$ & 4.66673(71) & 4.66695(68) & 4.67641(231) & 4.67065(160) & 4.67175(182) \\
$ 35 \times 10$ & 4.66990(24) & 4.67002(30) & 4.67634(44) & 4.67382(35) & 4.67362(43) \\
$ 40 \times 10$ & 4.67099(12) & 4.67098(14) & 4.67421(47) & 4.67299(25) & 4.67306(20) \\
$ 45 \times 10$ & 4.67166(13) & 4.67159(15) & 4.67371(22) & 4.67309(15) & 4.67318(19) \\
$ 50 \times 10$ & 4.67188(30) & 4.67199(31) & 4.67295(54) & 4.67253(52) & 4.67267(43) \\
$ 60 \times 10$ & 4.67244(19) & 4.67259(21) & 4.67326(29) & 4.67314(28) & 4.67318(25) \\
$ 80 \times 10$ & 4.67365(10) & 4.67364(11) & 4.67389(12) & 4.67386(11) & 4.67383(12) \\
\hline
$ 48 \times 12$ & 4.80980(17) & 4.80980(17) & 4.81303(48) & 4.81199(20) & 4.81205(26) \\
\hline
\end{tabular}
\caption{Critical couplings calculated using the fitting method explained in Sec.~\ref{sec:tc} and by the multihistogram analysis. }
\label{betacrittable}
\end{table}

\twocolumngrid

\bibliography{thermo}

\end{document}